\documentclass[a4paper,11pt]{article}
\usepackage{epsfig}
\usepackage{graphicx}
\usepackage{psfig}
\usepackage{epsfig}
\usepackage{rotating}
\usepackage{amssymb}
\usepackage{amsmath}
\usepackage{graphicx}
\usepackage{subfigure}

\def\beq{\begin{eqnarray}}
\def\eeq{\end{eqnarray}}

\begin{document}

\title{Behavior of heuristics and state space structure near SAT/UNSAT transition}
\author{John Ardelius$^1$ \& Erik Aurell$^{1,2}$}
\maketitle
\begin{center}
\begin{tabular}{ll}
1& Dept. of Physics, KTH -- Royal Institute of Technology\\
& AlbaNova - SCFAB SE-106 91~Stockholm, Sweden\\
2 & SICS Swedish Institute of Computer Science AB \\
& Kista, Sweden\\
\end{tabular}
\end{center}
\maketitle

\begin{abstract}
We study the behavior of ASAT, a heuristic for solving satisfiability
problems by stochastic local search near the SAT/UNSAT transition.
The heuristic is {\it focused}, i.e. only variables in unsatisfied
clauses are updated in each step, and is significantly
simpler, while similar to, {\it walksat} or {\it Focused Metropolis Search}.
We show that ASAT solves instances as large as $N=10^6$ 
in linear time, on average, 
up to $\alpha=4.21$ for random 3SAT. For $K$ higher than 3,
ASAT appears to solve instances at the 
Montanari-Ricci-Tersenghi-Parisi ``FRSB threshold'' 
$\alpha_{s}(K)$ in linear time, up to $K=7$.
\end{abstract}

\section{Introduction}
\label{s:introduction}
Satisfiability problems (SAT) appear critically in many
disciplines. Finding fast and reliable numerical methods for solving
them is crucial in industrial applications, such as scheduling,
or in verification. 

The random KSAT model, where each condition or clause has the same
arity $K$, and an instance is picked randomly, 
has been of interest both to theoretical computer science
and to statistical physics.
For $K\geq3$ KSAT belongs to the NP complete class of problems.
While a deterministic algorithm will always find
a solution if there is one, it presumably takes
a long time to solve a KSAT instance in worst case~\cite{Selman-P-NP-ref}.
However, the typical behavior can be different. Indeed, since the
beginning of the 90s it has been known that the
average running time of a deterministic algorithm
depends on $\alpha = \frac{M}{N}$ 
the number of clauses ($M$) per variable ($N$)~\cite{hardness}.
As $\alpha$ varies, a SAT/UNSAT transition is observed at a
critical value $\alpha_{cr}$: below the threshold, a typical
instance is satisfiable, while above $\alpha_{cr}$ it 
is typically unsatisfiable. 
The transition becomes sharper as $N$ increases~\cite{SAT_UNSAT}.
Deterministic algorithms
have (empirically) longest average run times close
to $\alpha_{cr}$~\cite{hardness,SAT_UNSAT,GomesSelman05}.

Stochastic search heuristics are not guaranteed to find
a solution, if there is one, but may on the other hand
greatly outperform a deterministic algorithm on a typical (solvable)
instance.
As $\alpha$ increases for given $K$ and $N$, the typical run-time of
a given heuristic increases, eventually diverging, at the latest
at $\alpha_{cr}$.
The most interesting behavior, if it can be established, is if
for some heuristic the typical run-time grows only linearly
in $N$ for sufficiently small $\alpha$, and also if
the distribution of run-times per variable gets more narrow as
$N$ increases. If so, run-time per variable is 
a self-averaging quantity.
We denote here the greatest 
such $\alpha$ for some heuristic $\alpha_{lin}$, 
the {\it linear-time transition} for that heuristic.

A benchmark stochastic search is 
Papadimitriou's RandomWalksat~\cite{RandomWalksat}: in every step
an unsatisfied clause is picked randomly, and then one
random variable in that clause is flipped. 
For that algorithm, rate equations and direct simulations 
indicate that  $\alpha_{lin}$ on 3SAT is approximately 
$2.7$~\cite{BARTHEL,SemerjianMonasson03}. 
Furthermore, for simple heuristics, such as straight-forward guessing
without back-tracking, rate equations and direct simulations
also show a nonzero $\alpha_{lin}$, albeit smaller~\cite{DeroulersMonasson04}.

A limitation of RandomWalksat is that it does not distinguish
between which variable in a clause to flip, or if flipping one increases
or decreases the number of unsatisfied clauses.
The {\it walksat}~\cite{walksat} algorithm mixes RandomWalksat moves
with greedy steps, by default in equal proportion.
Walksat has been known to be quite powerful on SAT problems, but
it was only shown quite recently to have a $\alpha_{lin}$ of 
$4.15$ on 3SAT~\cite{AurellGordonKirkpatrick04}.
In contrast to RandomWalksat, rate equations have not been set
up for walksat: the interleaving of random and greedy moves,
and the additional ``freebie'' move in the Selman-Kautz-Cohen
heuristic, makes that complicated.  
Alava, Orponen and Seitz showed that  
$\alpha_{lin}$ for walksat could be pushed to or beyond 4.20
by optimizing over the proportion of random and greedy moves~\cite{FMS}.
Furthermore, these authors showed that two other algorithms,
{\it Focused Metropolis Search (FMS)} and {\it Focused Record-to-Record Travel}
can be optimized to also have an apparent $\alpha_{lin}$ around 4.20.
FMS in particular is quite simple: a 
variable in an unsatisfied clause is flipped if that
decreases the number of unsatisfied clauses, and otherwise
flipped or not flipped by a probability exponential 
in that number. FMS does not have the freebie move of walksat,
but is still of comparable efficiency.

In this contribution we will introduce and study a heuristic ASAT,
for \textit{average SAT}, which is 
arguably yet simpler than FMS. In ASAT
a variable is flipped if this decreases the number of
unsatisfied clauses, as in FMS, and then flipped with
a \textit{constant} probability if the number of unsatisfied
clauses increases. ASAT is therefore sensitive to
the {\it widths} of local minima, but not directly to
the {\it heights} of the walls around local
minima.
  
The relevance of these studies lies along the following lines. 
First, more powerful search heuristics have a practical interest.
While we do no not present detailed comparisons in this
paper, let us state that ASAT generally runs somewhat faster
than optimized FMS, which in turn is somewhat faster than
optimized walksat with the Selman-Kautz-Cohen heuristic.
More importantly, since ASAT is simpler than FMS,
which in turn is simpler than walksat, one might hope
for a analytical treatment along the lines 
of~\cite{BARTHEL,SemerjianMonasson03}.
Second, from the theoretical side,
it is of interest if $\alpha_{lin}$ for ASAT and other heuristics
lie beyond $4.20$ on random 3SAT, since that lies beyond
two natural candidates for upper bounds on $\alpha_{lin}$,
known respectively as $\alpha_d$ (which is around $3.92$ for
random 3SAT) and $\alpha_s$ (around $4.15$).

The theoretical background of these two numbers
$\alpha_d$ and $\alpha_s$ can be briefly described as follows.
Within the cavity method, it has been shown that for
low enough $\alpha$ the set of solutions is connected,
while in the interval $[\alpha_d,\alpha_{cr}]$ the set
breaks up into ``clusters''~\cite{MZ,MPZ}. This
was recently rigorously confirmed 
for large enough $K$~\cite{MMZ-05,AchlioptasNaorPeres05}.
A satisfiability problem is equivalent to the problem
of finding a zero-energy ground-state in a statistical
mechanics model, where the ``energy'' is the number
of unsatisfied clauses. In the UNSAT regime, where an
instance is typically unsatisfiable, the ground-state
energy is typically larger than zero.
In the SAT regime, clusters of solutions are local minima,
which are also global minima. If these 
are accompanied by a much 
larger number of clusters with non-zero energy,
such clusters could act as traps to local search heuristics. 
The number of clusters of local minima at given energy
was computed by the cavity method for random 3SAT in~\cite{MZ,MPZ},
and for higher $K$ in~\cite{CLUSTER}, and does increase 
with energy.
One possible conjecture, already falsified 
in~\cite{AurellGordonKirkpatrick04},
would hence be that $\alpha_d$ is an upper bound on $\alpha_{lin}$. 
For KSAT a further phase transition takes place
at $\alpha_{s}$, inside the interval $[\alpha_d,\alpha_{cr}]$, 
giving rise to a hierarchical structure of 
clusters~\cite{MontanariParisiRicci-Tersinghi03}.
For random 3SAT, $\alpha_s$ is approximatively $4.15$.

It is not clear to the present authors what the relation, if any,
should be between $\alpha_{s}$ and $\alpha_{lin}$. 
Given however that $\alpha_{lin}$ for default walksat is
larger than $\alpha_d$ and very close to 
$\alpha_s$,
one might conjecture that $\alpha_s$ be an upper bound on
$\alpha_{lin}$.
The results of this paper and of~\cite{FMS} on the other hand
indicate that $\alpha_{lin}$ for optimized algorithms
is substantially larger than $\alpha_{s}$ for 3SAT.
We show that run-time
of ASAT is self-averaging at $\alpha=4.21$ on 3SAT up to instances
of one million variables, while it is not self-averaging
at $\alpha=4.25$. It is difficult to pin-point more precisely
the transition point in this interval. It is even more difficult
to compute the transition line of KSAT at $K$ larger than $3$ accurately,
as the memory requirements per variable grow as $K\alpha$. 
We present here data that at $\alpha_{s}(K)$,
as computed recently in~\cite{CLUSTER}, run-times of ASAT 
seem self-averaging up to $K=7$.
The time course of a solution is another quantity interest.
Below $\alpha_{lin}$, ASAT solves an instance in linear time,
similarly to RandomWalksat below its $\alpha_{lin}$. 
Above $\alpha_{lin}$, but of course on a satisfiable
instance, i.e. below $\alpha_{cr}$, ASAT typically solves
an instance by a slow process, ``sinking'' through several plateaus.
We show results from one such run, and we note that it
appears to be different from the ``solution by fluctuations''
proposed for RandomWalksat above its $\alpha_{lin}$~\cite{BARTHEL}.
To optimize ASAT we introduce a re-heating procedure, which
also sheds light on the effective landscape seen by the algorithm.

\section{The ASAT heuristic}
\label{s:asat}
ASAT is a focused heuristic, like RandomWalksat, walksat
and FMS, meaning it focuses on the
unsatisfied clauses at any given time,
and makes a trial moves only to neighboring states 
by flipping a variable that appears in at least one
unsatisfied clause. Variables that only appear in
satisfied clauses are never flipped.
In a sense, ASAT is perhaps the simplest extension
of RandomWalksat. For any trial move, it only
computes if that move will increase or decrease
the energy (number of unsatisfied clauses).
A move which increases the energy will be accepted
with fixed probability, independent on how
much the energy is changed, while a move that decrease
the energy will be accepted always. In pseudo-code, 
ASAT is hence

\begin{center}
\begin{figure}
\epsfig{figure=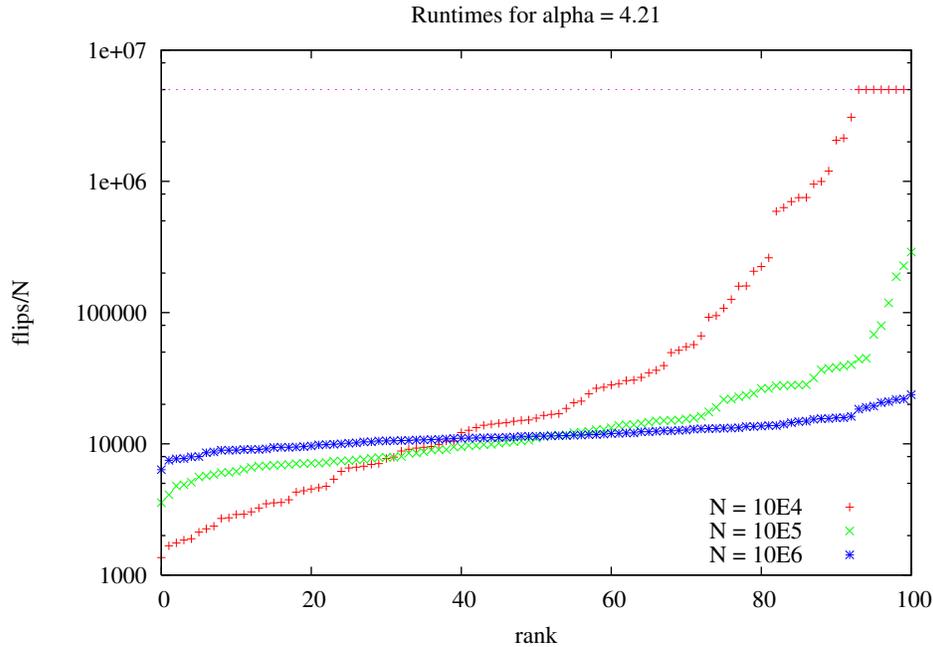,clip=}
\caption{Ranked logarithmic run-times per variable of 
ASAT, $p=0.21$, at $\alpha=4.21$ and values of 
$N$ from $10^4$ to $10^6$ on 3SAT. Note pivoting of the
distributions, as in \protect\cite{AurellGordonKirkpatrick04}.
Note that all runs were made with a cut-off of $5\cdot 10^{6}N$
flips. Out of one hundred, all instances at 
$N$ equal to $10^5$ and $10^6$ are solved within this time,
most instances at  $N=10^6$ taking close to $10^{10}$ flips.
For the smallest size, $N=10^4$, the spread is larger,
and about 10\% of the instances are not solved in $5\cdot 10^{10}$
flips, although the median is but a little more 
than $10^8$ flips.
}
\label{fig:a421rank}
\end{figure}
\end{center}

\begin{center}
\begin{figure}
\epsfig{figure=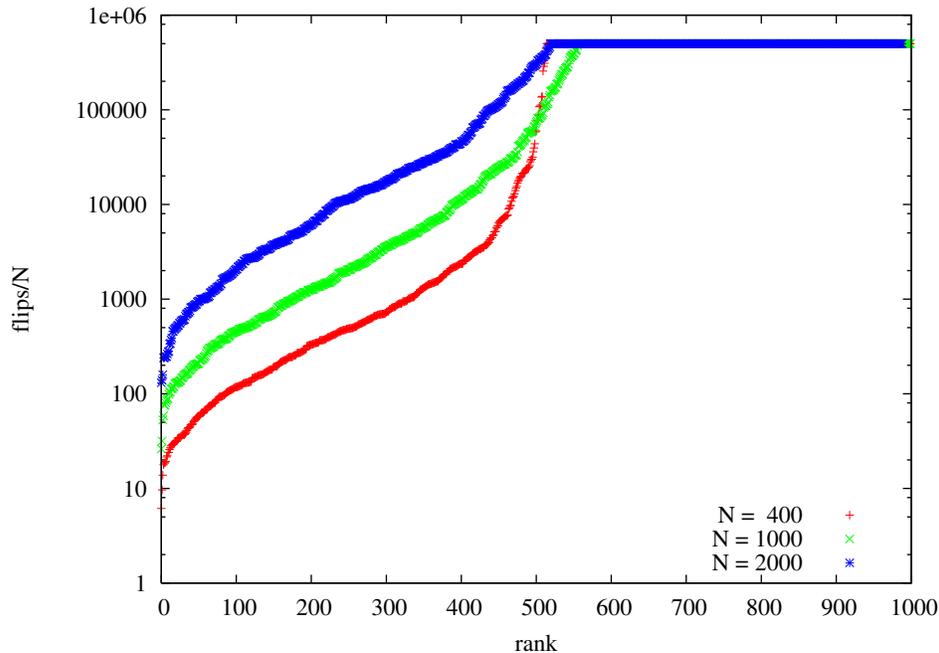,clip=}
\caption{Ranked logarithmic run-times per variable of 
ASAT, $p=0.21$, at $\alpha=4.25$ and values of 
$N$ from $10^4$ to $10^6$ on 3SAT. }
\label{fig:a425rank}
\end{figure}
\end{center}

\begin{verbatim}
s = initial random configuration

while t<t_max
       if F(s) = TRUE then EXIT
       at random pick a unsatisfied clause C
       at random pick a variable x in C

       x' = flip(x)
       s' = s(x -> x')

       if E(s') <= E(s) then flip x else
              flip x with probabililty p
\end{verbatim}
The ASAT algorithm is therefore characterized by the single
parameter $p$, which plays an analogous role to the
proportion of random and greedy moves in walksat
(a parameter also called $p$), and the noise parameter
$\eta$ of FMS. Optimization of $p$ in ASAT is discussed
below in section~\ref{s:structure-optimization}.
Fig.~\ref{fig:a421rank} shows a rank ordered plot of the
run-times for different system
sizes $N = 10^{4},10^{5},10^{6}$ at $\alpha=4.21$.
This and analogous data for several lower values of $\alpha$ (data not shown)
indicate that the run-time of ASAT is self-averaging at least this far.
As an example of lack of self-averaging, we show in
Fig.~~\ref{fig:a425rank} a rank ordered plot at  $\alpha=4.25$,
the conjectured end point of the SID (Survey Induced Decimation)
algorithm~\cite{MPZ,MZ}. Clearly here there is no sign of self-averaging. 
The plot in Fig.~\ref{fig:a421rank} is in a sense misleading,
as there are finite-size effects at smaller $N$. In other 
words, ``good pivoting'' and hence (numerically) 
convincing self-averaging
is only displayed from values of $N$ around $10^4$ and upwards.
Similar finite-size effects were noted by the authors of~\cite{FMS}
on optimized FMS (data unpublished) and by one of us on walksat, 
and are presumably a feature of the
constant-$\alpha$ ensemble. It seems that the onset of
good pivoting increases with $\alpha$, which makes 
a precise determination of $\alpha_{lin}$ difficult.
The largest instances that fit into memory of present generation
workstations is on the order of a million variables at 
$\alpha\sim\alpha_{cr}$: there are $\alpha N$ clauses, each one of them 
specified by $K$ integers,
which for $K=3$ and $\alpha\sim 4$ makes about $50\cdot N$ bytes.
Hence, for $\alpha$ larger than $4.21$ the interval 
between the onset of good pivoting and the largest 
instances that can be investigated is too small to draw 
a conclusion.
For higher values $K$ the problem quickly
becomes worse, since $\alpha_{cr}$ increases, roughly as $2^K$.
 
\begin{center}
\begin{figure}
\epsfig{figure=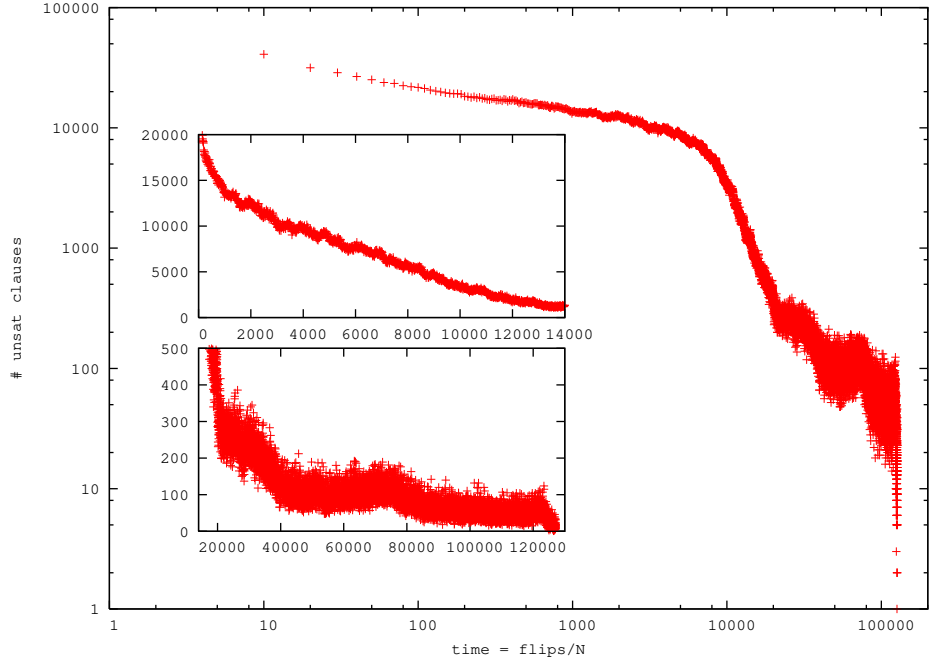,clip=}
\caption{Time course of solution of one instance by ASAT at
$\alpha=4.22$ and $N=10^6$. Note that ASAT solves this
instance, but only after $1.3*10^5$ time steps,
i.e. $1.3*10^{11}$ flips.
The main plot, in logarithmic coordinates, shows that the
solution proceeds in three stages. First, there is
decay on a time scale up to about $10^3$.
This process slows down, and is overtaken 
another process which last up to  
time about $2*10^4$. Finally, there is a very slow decrease to the solution.
Top insert shows a blow-up of the second stage in linear
coordinates. Bottom insert shows the final decrease, which
proceeds by plateaus, where the energy is approximately constant. 
In this run three plateaus can be discerned, with, approximately,
200, 50 and 15 unsatisfied variables, respectively.}
\label{fig:a422full}
\end{figure}
\end{center}

The solution process can be characterized by the 
fraction of unsatisfied clauses as a function of the
number of flips. Following~\cite{BARTHEL,SemerjianMonasson03,DeroulersMonasson04}
it is convenient to introduce a ``time'' as \textit{flips/N}.
Fig.~\ref{fig:a422full} shows the time course 
of a solution process at $\alpha=4.22$. One can clearly see three regimes, one
fast, one intermediate, and one quite slow. The fast regime, up
to time about ten thousand, is presumably
analogous to the Poissonian regime in RandomWalksat as 
studied by~\cite{BARTHEL,SemerjianMonasson03}.
The intermediate and slow regimes have, as far as we know,
not been shown on this problem previously. We note
that dynamics appears self-averaging in both the fast and
the intermediate regime;
perhaps hence both the fast and the intermediate regimes
will be amenable to analysis.
The slow regime proceeds by plateaus (long waiting periods).
While this qualitative behavior repeats itself from run to
run, the position and lengths of the plateaus do not.
In the slow regime, dynamics is hence not self-averaging.
We note that this solution mode, ``slowly sinking in the
energy landscape'',
seems qualitatively different from the ``solution by fluctuations''
found for RandomWalksat in~\cite{BARTHEL,SemerjianMonasson03}.

\begin{center} 
\begin{figure}
\epsfig{figure=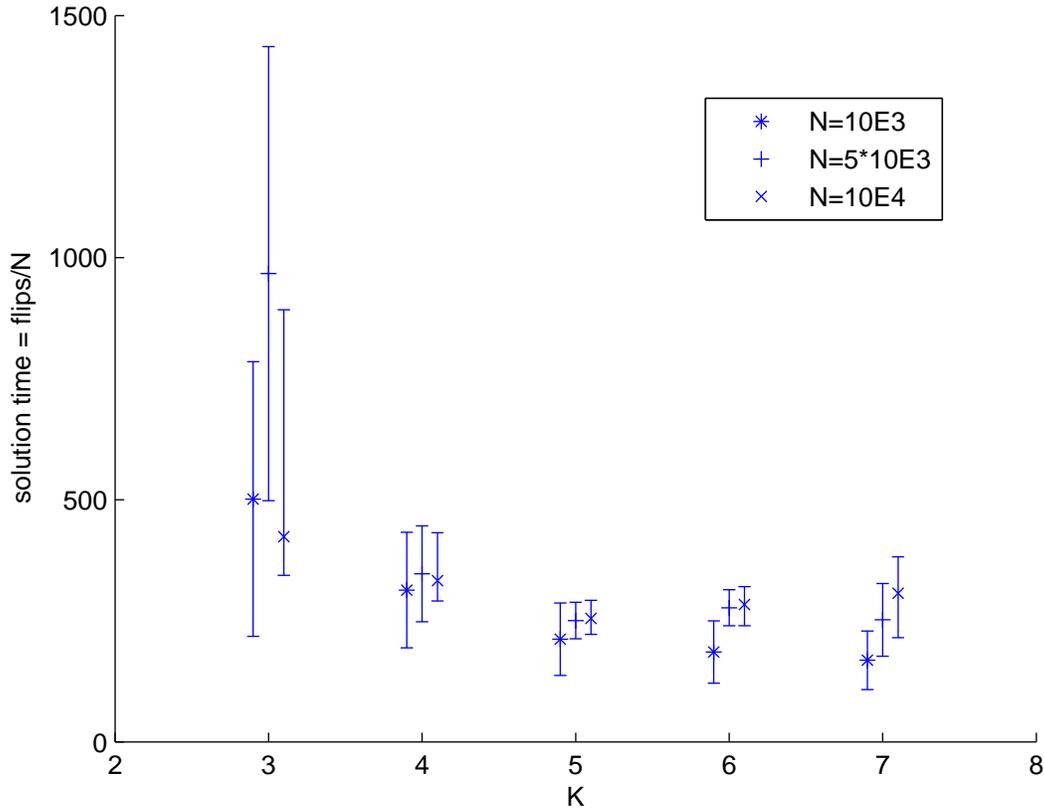,clip=}
\caption{
Median and quartile run times per variable at
$\alpha=\alpha_{s}(K)$, values taken from \protect\cite{CLUSTER}.
Figure indicates that ASAT solves these instances with
about equal computational cost per variable, for all $K$.
The parameter $p$ was found by procedure ASAT-HEAT 
(see main text) on each value of $K$ separately.
The values were $0.21$,  $0.118$, $0.068$, $0.045$ and
$0.032$ for $K$ from $3$ to $7$, respectively.
}
\label{fig:kscale}
\end{figure}
\end{center}

Finally, we have investigated ASAT for $K$ larger than 3,
albeit in less detail. While computational determination of
the threshold gets harder at higher values of $K$, one may
look for evidence that some given $\alpha$ is comfortably below
$\alpha_{lin}$. In Fig.~\ref{fig:kscale} we have looked
at $\alpha_s(K)$, which values were recently given
as $4.15$, $9.08$,  $17.8$, $33.6$ and $62.5$ for $K$ from
3 to 7, respectively~\cite{CLUSTER}.
The results are not entirely conclusive, but tend to support
that $\alpha_{lin}$ is greater than $\alpha_s$.  

\section{State space structure and parameter optimization}
\label{s:structure-optimization}
In this section we describe 
a method to optimize the value of the noise
parameter $p$. The method called \textit{simulated
heating of ASAT}, or ASAT-HEAT, is also,
as we will see, a useful tool when investigating the 
local barriers of random K-SAT. 

The idea is that there is a trade-off to be made between
an algorithm getting out of local minima, and
efficiently exploring the bottom of a local minimum.
Hence, the premise is that solutions are found at the
bottom of \textit{some} local minima, which are not otherwise distinguished.
After completion of this work, we became aware of a 
related idea, ``optimization at the ergodic edge'',
has recently been considered by Boettcher and 
Frank~\cite{BoettcherFrank05}, and also,
to optimize the Record-to-Record-Travel algorithm,
by Jia, Moore and Selman~\cite{JiaMooreSelman05}.

\begin{center}
\begin{figure}
\epsfig{figure=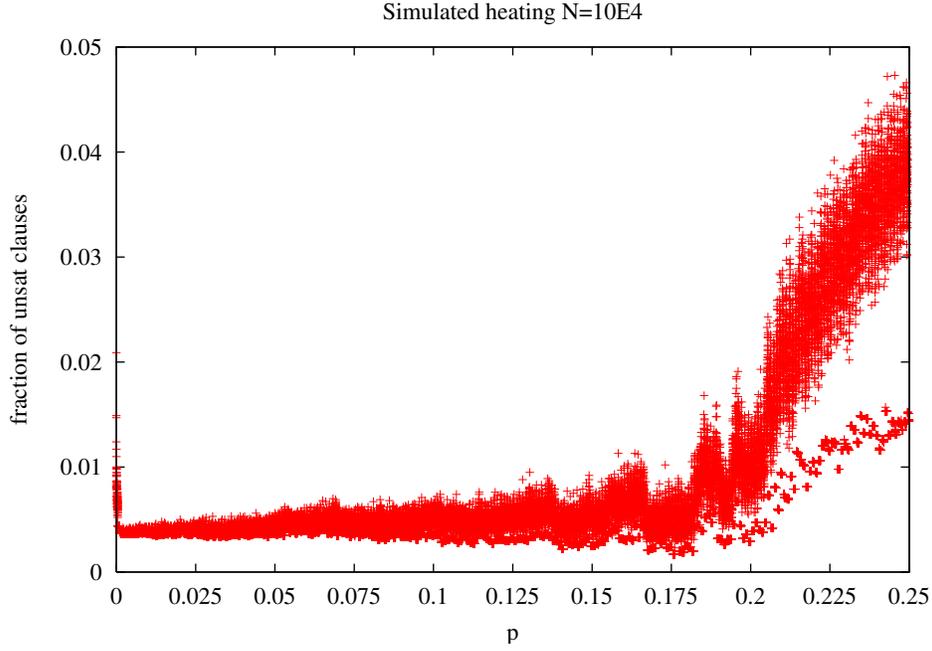,clip=}
\caption{Escape from meta-stable state at $\alpha=4.27$ and
$N=10^4$
while increasing parameter $p$ in procedure ASAT-HEAT (see main text).
Note that up to a critical value $p_{cr}$ fluctuations are small and
the energy decreases slowly. At higher values of $p$, fluctuations are
larger, and energy increases with $p$.
Figure shows every $N$th time point during the heating periods
($p$ greater than zero) and the zero-temperature
quenches, at each value of $p$, respectively.}
\label{fig:heat1}
\end{figure}
\end{center}

In the context of ASAT we look for the value $p$ such that
the algorithm does not get stuck, while still exploring 
the bottom of mimina where it finds itself. That is
done by an interleaved process, where the algorithm 
alternatively runs with some non-zero $p$ (to explore phase
space, and get out of minima), and alternatively freezes
at zero $p$ (to find the bottom of the mininum it is moving in).
In ASAT-HEAT the value of $p$ is raised incrementally in the following steps:
\begin{itemize}
  \item Generate an problem instance of size $N$ 
        close to $\alpha_{cr}$.
  \item Run the heuristic with a low value of the parameter $p$
  and let the system equilibrate for some time $\tau>>1$.
  We have used $\tau$ equal to one thousand, \textit{i.e.}
  $10^3 N$ flips.
  \item Do a \textit{zero temperature quench}, that is set the noise
  parameter to zero. The heuristic will then find a near local
  optimum by greedy search.
  \item Reset the parameter $p$ to its previous value, 
   increase it by a small amount $\Delta p$, and let
   the system again equilibrate for a time $\tau$.
  \item Iterate heating and quench.
\end{itemize}
Fig.~\ref{fig:heat1} shows ASAT-HEAT for a system at $K=3$, $N=10^{4}$ and
$\alpha=\alpha_{c}=4.27$.
Up to $p_{cr}\sim 0.21$ the variations during the heating periods
increase with $p$, while the lowest energies reached after 
the zero-temperature quench trend downward. 
Furthermore, the lowest energies seen during the heating periods
are quite similar to the lowest energies after quench.
This suggests that the algorithm is here in contact with the
local structure, i.e. that the algorithm visits the local minima
also during the heating periods.
The downward trend of the lowest energies after quench indicates
that the algorithm over time explores a larger set of minima,
reaching lower energies, compare the time course of a simulation
as in Fig.~\ref{fig:a422full}.
After $p_{cr}\sim 0.21$, the variations during the heating periods
are larger. The energies reached after quench are clearly lower
than the ranges seen during the heating periods, indicating that
the algorithm has now lost contact with the local structure.
The energies reached after quench on the other hand
increase with $p$, indicating that the algorithm now moves
in higher-lying and more numerous local minima. 
We have found that ASAT works most efficiently 
using values of $p$ at or slightly above $p_{cr}$.

\begin{center}
\begin{figure}
\epsfig{figure=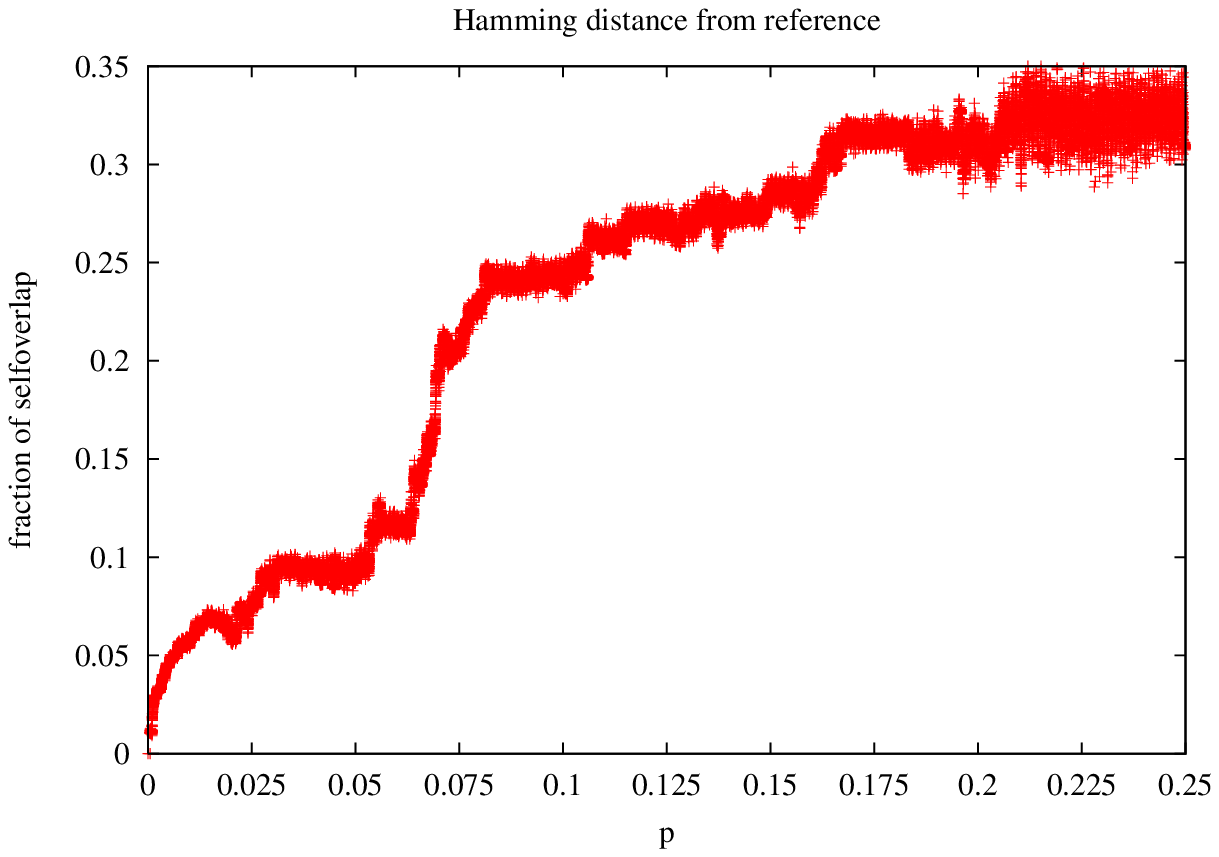,clip=}
\caption{Same run as in Fig.~\protect\ref{fig:heat1}.
Hamming distance from initial meta-stable state
increases with $p$ by plateaus up to the critical value,
at which is remains stationary. Fluctuations are relatively
larger above the critical $p$.}
\label{fig:hamming2}
\end{figure}
\end{center}

To investigate how the algorithm moves in state space, 
we compute the Hamming distance between the state
at some time (some value of $p$ in ASAT-HEAT), and a reference
state at the initial value of $p$.
Fig.~\ref{fig:hamming2} shows the variation in Hamming distance for 
system size $N=10^{4}$, which exhibits a step-wise increase 
by plateaus. Fluctuations in Hamming distance 
are small below $p_{cr}$, and increase rapidly above $p_{cr}$.
We have further investigated how $p_{cr}$
varies with $N$ by running simulations at ten times the original
system size. The overall behavior is similar and the transition
seems to take place at or close to the same value of $p$. This 
suggests that the local small scale structure does not depend
or depends only weakly on $N$.
The value of $p_{cr}$ found by ASAT-HEAT decreases with
$K$; the values for $K$ from $3$ to $7$ are given in caption
to Fig.~\ref{fig:kscale}

\section{Discussion}
\label{s:discussion}
We have in this work presented a new heuristic for satisfiability
problems, called ASAT. The heuristic has been implemented
as an additional option in walksat
(walksat distribution format~\cite{walksat-WWW}), and is available upon
request. One main interest of ASAT is that it is competitive with
walksat and Focused Metropolis Search (FMS) on random KSAT
problems close to the SAT/UNSAT transition, while being simpler,
and hence (hopefully) amenable to analytic investigations.
We hope to return to this question in a future contribution. 

We have shown that ASAT has typical runtime
linear in $N$ up to $\alpha=4.21$ on 3SAT, up to the largest
instances that can be studied ($N\sim 10^6$). We can 
not exclude that ASAT will eventually not look
linear on even higher values of $N$, but we see no sign
of such divergence. Finite-size effects make it difficult
to show if or if not ASAT is linear or not beyond $\alpha=4.21$.
This means, that to the best of our estimate,
the linearity threshold for ASAT, $\alpha_{lin}$,
is for random, 3SAT larger than both the clustering transition
($\alpha_d=3.92$) and the ``FRSB threshold'' ($\alpha_s=4.15$).
We have studied ASAT at larger $K$, and showed that
probably $\alpha_{lin}$ is likely to be larger than $\alpha_s$ there also.

A parameter optimization technique, ASAT-HEAT, has been
introduced. This allows for a determination of an optimal
parameter value of the algorithm, and can be considered
an alternative to the extensive simulations at many values
of $N$, $\alpha$ and one algorithm parameter used in~\cite{FMS}.

While physical intuition suggests that local heuristics will have
difficulties where many meta-stable states appear (at $\alpha_d$, or 
perhaps, more properly, at $\alpha_s$) this does not seem to be the
case. RandomWalksat and very simple heuristics have 
difficulties far below $\alpha_d$, while 
ASAT and other heuristics seem to work linearly 
beyond $\alpha_s$. 
Let us therefore end by stating the differences between 
stochastic local search heuristics to find satisfying assignments in random KSAT
and a physical process of random walk in a corresponding energy landscape.
First, RandomWalksat,
walksat, FMS and ASAT are focused: these algorithms correspond
to non-equilibrium dynamics  without detailed balance~\cite{SemerjianMonasson03}. 
Conservation of any cluster structure at all under non-equilibrium 
perturbations has apparently been a moot point in some spin glass 
models~\cite{CrisantiSompolinsky87,CugliandoloKurchanLeDoussalPeliti96}.
Second, while FMS is similar to a random walk in the energy landscape,
in the sense that the dynamics directly depends on the local energy, 
walksat and ASAT and obviously RandomWalksat do not.
In walksat with Selman-Kautz-Cohen heuristic, decisions are based
on the change in ``breakclause'' which is not the same as the energy change, 
while in ASAT decisions are based on whether the energy increases
or decreases at all. 
Therefore, finally, 
while numerical simulations cannot rule out that
\textit{e.g.} ASAT will run into trouble beyond $\alpha_s$ 
on instances larger than the ones we have studied,
we are not sure if it necessarily has to.
Theoretical predictions on what $N$ one would expect to see
nonlinear behavior at what $\alpha$ would be most helpful
to guide numerical experiments on this issue.

\section*{Acknowledgements}
This work was
supported by the
Swedish Research Council (E.A).
We thank Supriya Krishnamurthy for numerous discussions and 
a close reading of the manuscript. We thank
Mikko Alava, Jean-Philippe Bouchaud,
Scott Kirkpatrick, Pekka Orponen, Giorgio Parisi, Sakari Seitz,
Guilhem Semerjian and Riccardo Zecchina 
for discussions and critical remarks.

\bibliographystyle{unsrt}
\bibliography{citations}
\end{document}